\documentclass[twocolumn,superscriptaddress,showpacs,PRB,amsmath,amssymb]{revtex4} 
\usepackage{graphicx} 
\usepackage{dcolumn}
\usepackage{bm}
\usepackage{epsfig}

\begin{document}

\title{Quantum Oscillations in the High Pressure Metallic State of Ca$_2$RuO$_4$}
\author{Swee K. Goh}
\author{P. L. Alireza}
\author{Lina E. Klintberg}
\affiliation{Cavendish Laboratory, University of Cambridge, J.J. Thomson Ave, Cambridge CB3 0HE, United Kingdom }
\author{T. Murphy}
\affiliation{Florida State University, National High Magnet Field Laboratory, Tallahassee, FL 32306 USA}
\author{F. Nakamura}
\affiliation{Department of Quantum Matter, ADSM, Hiroshima University, Higashi-Hiroshima 739-8530, Japan}
\author{David J. Singh}
\affiliation{Materials Science and Technology Division, Oak Ridge National Laboratory, Oak Ridge, Tennessee 37831-6056, USA}
\author{Mike Sutherland}
\affiliation{Cavendish Laboratory, University of Cambridge, J.J. Thomson Ave, Cambridge CB3 0HE, United Kingdom }

\begin{abstract}
We present evidence for quantum oscillations in the pressure-induced metallic state of the 4$d$ layered perovskite Ca$_2$RuO$_4$. A complicated oscillation spectrum is observed, which is both temperature and field dependent, with unusually light cyclotron masses in the range of $m^*/m_e$ $\sim$ 0.6 -- 3, suggesting that the pressure-induced metallic state is a weakly correlated Fermi liquid. We compare our observations to band structure calculations within the local spin density approximation, and conclude that some features of the spectrum are a result of non-linear spin splitting effects.

 \end{abstract}

\pacs{71.18.+y, 74.70.Pq, 71.30.+h}
\maketitle
The Mott transition from a metallic state with mobile charge carriers, to a state where electrons are held in place through strong Couloumb interactions is a dramatic manifestation of many body physics. Theoretical and experimental studies of the Mott transition have a long and fascinating history \cite{Imada98}, motivated in part by materials such as the manganites and high $T_c$ cuprates. A complete microscopic description of the Mott transition, particularly in complex multiband metals, is however still lacking.

One particularly interesting Mott insulating compound is the layered ruthenate Ca$_2$RuO$_4$, which possesses a rich phase diagram when tuned by applying hydrostatic pressure or by chemical substitution. At ambient conditions Ca$_2$RuO$_4$ is a spin-one, nearest neighbor ordered antiferromagnetic Mott insulator with $T_N$ $\sim$ 110~K \cite{Nakatsuji97,Cao97}. It has a layered perovskite structure (Pbca) containing RuO$_6$ octahedra that are tilted through the $ab$ planes, rotated about the $c$-axis and compressed \cite{Braden98}. The replacement of Ca with Sr in Ca$_{2-x}$Sr$_x$RuO$_4$ has a pronounced effect on these structural distortions, which ultimately determine the magnetic and electronic states of the material. For a Ca concentration in the region $0.5 > x > 0.2$ the low temperature structure remains Pbca but with an increased $c$-axis lattice parameter. The system becomes metallic without long range magnetic order, however there is strong competition between a tendency towards ferromagnetism and nesting induced antiferromagnetism, the details of which are very sensitive to structure.  

As $x$ is further increased, the tilt of the octahedra is reduced, and is completely removed by $x$ = 0.5. Here the low temperature structure changes to I4$_1$/acd, and the rotational distortions show stacking sequences along the $c$-axis \cite{Friedt01}. At this concentration the susceptibility becomes Curie-like with an S = 1/2 moment per Ru-ion. The evolution of the ground state through this region of the phase diagram has led to the proposal that for $x > 0.5$, some $d$-electrons are localized, while others are itinerant -- a selective Mott transition \cite{Anisimov02}. At higher Sr concentrations the RuO$_6$ octahedra rotations are eventually suppressed and the crystal structure is I4/mmm, with the end member Sr$_2$RuO$_4$ being a well-known unconventional superconductor with no structural distortions \cite{Mackenzie03,Maeno12}.

The application of hydrostatic pressure to Ca$_2$RuO$_4$ offers a clean way to tune the properties of the system without the concomitant introduction of disorder from chemical substitution. Pressure produces a phase diagram at low temperatures that is similar to that of Ca$_{2-x}$Sr$_x$RuO$_4$, with a few important distinctions. At a modest pressure of 0.6~GPa a Mott transition occurs, driven by an abrupt and discontinuous lattice distortion. The resulting phase is a complicated mixed state of metallic, ferromagnetic regions coexisting with the antiferromagnetic Mott insulating state \cite{Nakamura02,Nakamura09}. Both phases share the Pbca crystal structures, but have markedly different $c$-axis lattice constants, which is expected to lead to strongly discontinuous behavior in this region of the phase diagram. For $p > 2.0$~GPa the ferromagnetic phase is dominant and $T_C$ jumps from 9~K to 20~K, yielding a highly anisotropic 2D metallic phase with a low residual resistivity. In this phase the structure remains Pbca, but without the stacking sequences seen in the chemically substituted system. The tilt of the RuO$_6$ octahedra is suppressed as pressure is increased, disappearing by $p = 5.5$~GPa \cite{Steffens05}. Eventually ferromagnetic order is also destroyed for $p > 7.5$~GPa, where the system undergoes a continuous structural transition to the orthorhombic Bbcm structure \cite{Steffens05}. At very high pressures on the order of $p \sim$ 9.0~GPa \cite{Alireza10} and above, the system enters a superconducting state with $T_c$ $\sim$ 400~mK.

The quasi-2D, pressure-induced metallic phase in the vicinity of the Mott transition in Ca$_2$RuO$_4$ is the subject of our study. To date, only limited transport and susceptibility studies have been reported \cite{Nakamura02,Alireza10} in this phase, and we here present measurements of quantum oscillations at high magnetic fields that show long-lived quasiparticles with light cyclotron masses arising from a homogeneous metallic phase. Band structure calculations using lattice parameters measured under pressure reveal a complicated Fermi surface formed through band folding resulting from structural distortions, and spin splitting arising from the presence of ferromagnetic order. Our results can be partially, but not fully explained by the geometry of the Fermi surface, and suggest the presence of non-linear spin splitting effects arising from proximity to a metamagnetic transition.

\section{Experimental}
Our study used high quality single crystal samples of Ca$_2$RuO$_4$ grown by a floating zone method with a RuO$_2$ self-flux \cite{Nakatsuji97,Nakamura02}. Electron-probe microanalysis showed no evidence for secondary phases or inclusions, and the samples were shown to have essentially stoichiometric oxygen concentration. Crystals were chosen from a batch having $\rho_0 \sim 1.6 - 3.0~\mu\Omega$cm in the metallic state at high pressures.

We observed quantum oscillations in measurements of the magnetic susceptibility of our samples via the de Haas -- van Alphen (dHvA) effect. To detect small changes in susceptibility we used a novel Moissanite anvil pressure cell design which incorporates a 140 turn primary drive coil inside the cell body. The pressure region between the anvils contained a 10 turn pickup coil made from 12~$\mu$m diameter copper wire, in which samples of typical size $\sim$ 70 $\mu$m $\times$ 180 $\mu$m $\times$ 40~$\mu$m rested \cite{Alireza03,Goh08}. For typical runs we used an excitation frequency of 10~kHz, and detected the pickup voltage using a lockin amplifier and transformers. To ensure as hydrostatic conditions as possible, we used argon as a pressure transmitting medium. Pressure was measured using shifts in the ruby fluorescence spectrum from small ruby chips embedded within the pressure region. 

Our quantum oscillation measurements were carried out in two laboratories. First, we used an 18~T superconducting magnet with a dilution refrigerator at the University of Cambridge. Followup experiments were conducted using a 31~T DC magnet with a $^3$He insert at the National High Magnetic Field Laboratory (NHMFL) in Tallahassee. Samples were oriented such that the magnetic field was applied along the $c$-axis, with cyclotron orbits occurring within the plane.

\section{Low-field Results}

       \begin{figure} \centering
             \includegraphics[bb=80 20 500 790,scale=0.45]{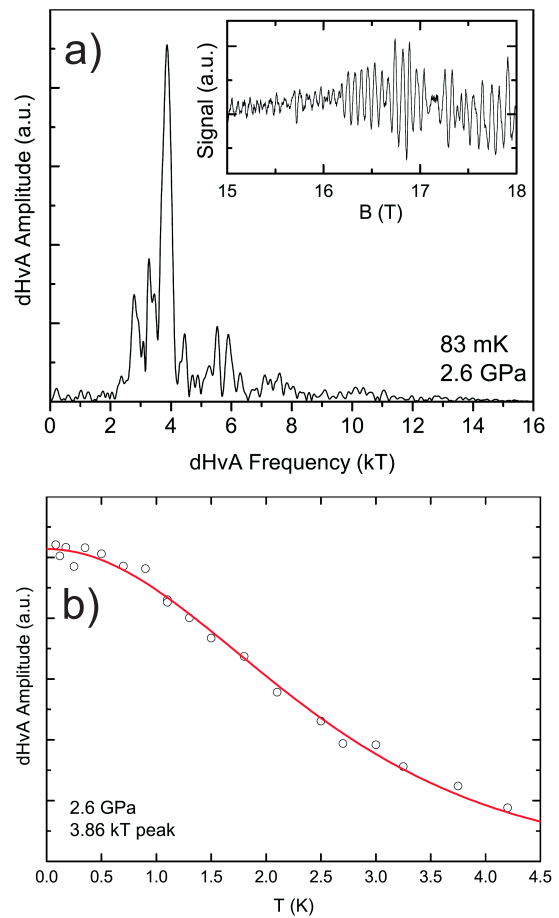}
              \caption{\label{fig:1} Top: Fourier transform of the magnetic susceptibility data in the inset using a window between 15-18 T. A strong peak is seen at 3.86~kT, and less well resolved peaks at lower frequencies. Top Inset: AC susceptibility of Ca$_2$RuO$_4$ at $p$ = 2.6~GPa and $T$ = 83~mK, with $B \parallel c$. Bottom: Temperature dependence of the amplitude of the 3.86~kT peak. The red line is a fit to Lifshitz-Kosevich damping factor in Eq. \ref{eq:LK}.}
              
      \end{figure}
   
Figure \ref{fig:1} shows the results for a typical run on our 18~T cryomagnetic system. The inset of the upper panel shows the pickup voltage from the secondary coil after a Savitzky-Golay method was used to subtract the field dependent background. Clear evidence of oscillations are seen, with a field dependent amplitude envelope that is strongly modulated. This is reflected by the closely spaced family of peaks in the Fourier transform of the data, shown in the main panel of Fig. \ref{fig:1}a. The experiment was performed at a base temperature of 83~mK, which is limited due to self heating of the cell from the excitation current. Given the constraints on the excitation field size $b_0$ imposed by the small dimensions of the coil, it is favorable to detect the fundamental frequency instead of higher harmonics as is often done in  dHvA experiments. For this configuration, the signal amplitude for a quantum oscillation with frequency $f$ is proportional to $J_1(\lambda)$, where $\lambda = 2\pi f b_{0}/B^2$, and for small $\lambda$, $J_1(\lambda) \propto \lambda$. If heating is significant, the signal amplitude will deviate from this linearity. This allows us to pick a modulation field to minimize the heating effect, and we take the sample temperature to be the same as that of the mixing chamber \cite{Goh08}. The applied pressure was measured to be 2.6 GPa, within the metallic phase as determined by Nakamura and coworkers \cite{Nakamura02}. In the context of the doping phase diagram of Ca$_{2-x}$Sr$_x$RuO$_4$, a sample at this pressure is most similar in structure to compounds with a doping in the range $x$ = 0.2 -- 0.5.
      
The dHvA frequency $f$ (in Tesla) is related to the extremal cross-sectional area of the Fermi surface $A_{ext}$, through the Onsager relation $f = \hbar A_{ext}/2\pi e$ \cite{Shoenberg84}. In a simple non-magnetic system with well separated oscillation frequencies, the field and temperature dependence of the amplitude of the oscillations are expected to be governed by the Lifshitz-Kosevich damping factors $R_D$ and $R_T$.

\begin{equation}
\label{eq:BD}
 R_D = \textrm{exp}(-B_D/B), B_D = \frac{\hbar C_F}{2e l}
\end{equation}

\begin{equation}
\label{eq:LK}
R_T=\frac{X}{\textrm{sinh} X}, X = 14.693 \frac{pTm^*}{B}
\end{equation}

\noindent where $l$ is the mean free path and $C_F$ is the circumference of the orbit in $k$-space. In Fig. \ref{fig:1}a, the presence of several closely spaced oscillation frequencies in the Fourier spectrum makes the field and temperature dependence of the amplitudes governed by Eqns. \ref{eq:BD}-\ref{eq:LK} challenging to analyze quantitatively. The dominant peak in the spectrum is centered at $f$=3.86~kT, with minor peaks discernible at 3.28~kT and 2.80~kT. There is some spectral weight at higher frequencies, but these peaks are close to the noise limit. To confirm that the oscillations we observed arise from the Landau quantization of cyclotron orbits in the usual way, we track the amplitude of these peaks as a function of temperature, with the 3.86~kT shown as an example in Fig. \ref{fig:1}b. The temperature dependence follows very closely that expected from Eqn. \ref{eq:LK}, with a cyclotron mass of $m^*$ = (1.04 $\pm$ 0.02)$m_e$. Similar fits to the 2.80~kT and 3.28~kT peaks yield  masses of (0.56 $\pm$ 0.04)$m_e$ and (0.78 $\pm$ 0.04)$m_e$ respectively. We take this as strong confirmation that the pressure induced metallic state in Ca$_2$RuO$_4$ is a conventional Fermi liquid, in the sense that it supports long-lived quasiparticle excitations. 

Previously reported resistivity measurements suggest that this metallic state is highly anisotropic.  At $p = 2.0$~GPa in Ca$_2$RuO$_4$, the temperature dependence of $\rho_c$ is non-metallic at temperatures $T>25$~K, arising from incoherent processes. However below this temperature $\rho_{c}$ is metallic, but with $\rho_{c}/ \rho_{ab}$ $\sim$ 1000 at low temperatures \cite{Nakamura02}, implying that the orbits we observe are likely to be from quasi-two dimensional Fermi surface sheets.  

We can also infer from the presence of these oscillations that the metallic state at $p = 2.6$~GPa must be fairly homogeneous. Measurements of Raman scattering at low temperatures in Ca$_2$RuO$_4$ have revealed a two-magnon peak that persists to pressures higher than 1.0~GPa. The existence of this peak is linked to the superexchange of spins on neighboring Ru atoms, and is thus related to the presence of short range antiferromagnetic correlations \cite{Snow02}. This has been interpreted as evidence for an inhomogeneous metallic state, with phase separated islands of AF ordered insulator. A similar conclusion was reached via diffraction studies \cite{Steffens05}, where an equal volume fraction of insulating and metallic phases was seen at $p = 1.0$~GPa. 

We can place a limit on the size of any such inhomogeneity in our measurements by estimating the diameter of the cyclotron orbits, which must occur inside metallic regions. We can compute the radius of the real space orbit $r_c$ = $k_F \hbar/eB$, using an average $k_F$ estimated from Onsager's relation. At 15~T, the lowest field where we recorded our signal, the cyclotron radius associated with the 3.86~kT orbit is $r_c \sim$ 150~nm, which puts a lower limit on the size of the metallic regions. The relatively large value of $r_c$, coupled with a low residual resistivity of this sample suggests that any widespread phase separation, if it exists, is removed by $p$ = 2.6~GPa and the sample is in a homogeneous metallic state.

       \begin{figure} \centering
           \includegraphics[bb=70 260 500 640,scale=0.54]{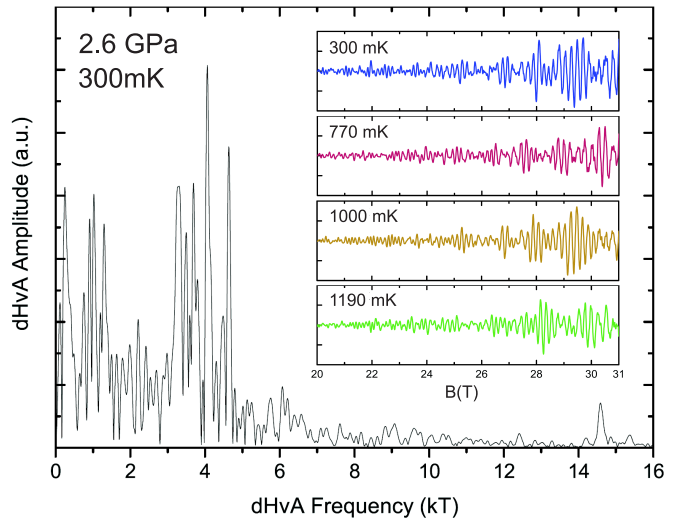}
              \caption{\label{fig:2}Main: Fourier transform of the high field quantum oscillation spectrum in Ca$_2$RuO$_4$ at $p = 2.6$~GPa, and $T = 300$~mK. The transform range was between 20 -- 31~T. Inset: The raw susceptibility data for the high field runs shown at several temperatures. }
              
      \end{figure}

\section{High-field Results}

In a simple metal, increasing the field range of quantum oscillation measurements has the effect of sharpening the resolution of oscillation frequencies. In order to further resolve the peaks observed in Fig. \ref{fig:1}a, we extended the field range of our measurements by using our pressure cell with 31~T resistive magnets at the NHMFL. The inset of Fig. \ref{fig:2} shows examples of our data taken at several temperatures, on the same sample, in the same cell at $p = 2.6$ ~GPa. We used similar excitation parameters to the low field measurements. Again, clear oscillations are observed, but the field dependence of the amplitude of the oscillations remains complicated. This is confirmed by Fourier transform of this data set seen in the main panel of the figure. Roughly speaking, the peaks can be grouped into three categories -- a cluster of peaks around $f \sim$ 1~kT, a cluster of peaks of stronger amplitude around $f \sim$ 4~kT, and a single, well defined peak at $f$ = 14.6~kT which was not present in the low field data. The data shown was Fourier transformed using a window of 20 -- 31~T, however it is important to note that the fine structure of the spectrum changes with field, with various peaks in the spectrum both shifting in amplitude and in frequency.

\begin{figure} \centering
             \includegraphics[bb=70 210 550 670,scale=0.45]{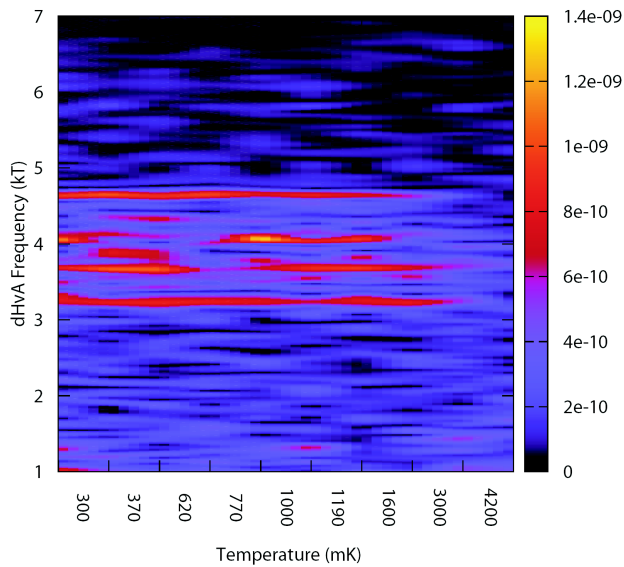}
              \caption{\label{fig:3}An intensity plot showing the Fourier transform spectral weight as a function of temperature for frequencies near $f$ = 4~kT. Increasing the temperature causes interference between frequencies which may suppress or enhance their amplitude leading to a non-monotonic temperature dependence.}
                    \end{figure}

Extracting information from the temperature dependence of the amplitude of these high field measurements is complicated -- as in the low field case, closely spaced frequencies cause interference that invalidates the straightforward use of Eq. \ref{eq:LK}. Small changes in temperature were observed to cause a pronounced change in the oscillation spectrum as the relative contribution of some frequencies changed and produced maxima or minima in the Fourier spectrum.  To illustrate this point, figure \ref{fig:3} shows an intensity plot demonstrating the effects of temperature on the amplitude of the oscillations in the 4~kT cluster of peaks. For some frequencies near 4~kT, a monotonic temperature dependence is observed and a good fit to Eq.\ref{eq:LK} is achieved \cite{endnote1}. For others, interference effects prohibit such an analysis, including all of those peaks in the 1~kT cluster. In contrast, the 14.6~kT  frequency shows a well defined temperature dependence that is easily fit, yielding a relatively heavy mass of $m^*$ = (3.63 $\pm$ 0.26)$m_e$. A summary of the fitted masses, where fitting is possible, is shown in Table 1.

These masses are suprisingly light in comparison with other layered ruthenate materials. The are much less than the cyclotron masses in Sr$_2$RuO$_4$ for instance, which lie in the range 3 -- 16~$m_e$ \cite{Mackenzie96}, and less than those measured by dHvA in Sr$_3$Ru$_2$O$_7$ (5 -- 10~$m_e$) \cite{Mercure10}. Similarly, Shubnikov-de Haas measurements in the ruthenate ferromagnet SrRuO$_3$ revealed frequencies close to those observed in the present study, (1.5~kT and 3.5~kT)  \cite{Mackenzie98,Alexander05}, with masses again considerably larger (4.5 and 6.1~$m_e$ respectively).

\begin{table}[h]
\caption{Effective masses in Ca$_2$RuO$_4$ at $p = 2.6$~GPa}
\vspace{10pt}
\centering
\begin{tabular}{ c | c | c  }

  $f$ (kT) & $m^*/m_e$ & Fitting range \\
  \hline
  2.80 & 0.56 $\pm$ 0.04 & 15 -- 18~T \\
  3.28 & 0.78 $\pm$ 0.04 & 15 -- 18~T \\
  3.86 & 1.04 $\pm$ 0.02 & 15 -- 18~T \\
    
  \hline
  3.24 & 0.58 $\pm$ 0.18 & 20 -- 31~T \\
  3.69 & 0.67 $\pm$ 0.18 & 20 -- 31~T \\
  4.64 & 1.21 $\pm$ 0.12 & 20 -- 31~T \\
  14.60 & 3.63 $\pm$ 0.26 & 20 -- 31~T \\
\hline  
\end{tabular}
\end{table}

Finally, we note that we confirmed our results by repeating the same set of high and low field experiments at a slightly higher pressure ($p = 3.2$~GPa) on a second sample, recovering similar frequencies, masses, and strong temperature and magnetic field dependences of the oscillation spectrum (c.f. Fig. \ref{fig:6}).
 
\section{Discussion}

The unusual field dependence of the oscillation spectrum seen in our high field experiments clearly presents a challenge for a detailed comparison with theoretical models. We propose two scenarios in which this complex spectrum might arise. First, it could be the result of a complicated band structure giving rise to closely spaced orbits as a result of Fermi-surface reconstruction from lattice structural modulations. Second, the multiple frequencies might arise as a result of a non-linear field dependent magnetization, which is known to produce subtle complications in the analysis of dHvA data.  We deal with these scenarios in detail below. A third possible explanation might be more prosaic in nature -- an unaccounted for effect due to our experimental setup. We find this scenario unlikely however, as previous measurements using the same setup were able to resolve the expected spectrum of oscillations in Sr$_2$RuO$_4$ \cite{Goh08}. We can also rule out any effects from the pressure medium. Argon is thought to be very hydrostatic at these pressures \cite{Tateiwa09}, and any non-hydrostaticity would lead to a loss of oscillation intensity rather than a peak splitting effect. A comparative study on Sr$_2$RuO$_4$ using a less hydrostatic medium (silicone oil) did not see a significantly altered oscillation spectrum.

\subsection{Band structure}

Density-functional theory (DFT) band structure calculations at ambient pressure in paramagnetic Ca$_2$RuO$_4$ predict a metal. The bands nearest to the Fermi level are formed primarily from Ru 4d $t_{2g}$ and O $2p$ states, with a bandwidth that is narrower than Sr$_2$RuO$_4$ due to tilting of the RuO$_6$ octahedra that changes the Ru-O-Ru bond angle \cite{Woods00}. Like Sr$_2$RuO$_4$, bands are labeled by $\alpha$ and $\beta$ (formed mainly from the $d_{yz}$ and $d_{zx}$ states) and $\gamma$, arising from the $d_{xy}$ states. A strong competition between ferromagnetic and antiferromagnetic correlations is also predicted, with an incommensurate antiferromagnetic order arising from strong nesting between the $\alpha$ and $\beta$ bands, and ferromagnetism arising from a van-Hove singularity in the $\gamma$ band \cite{Mazin97,Mazin99,Fang01}. The observed insulating behavior of Ca$_2$RuO$_4$ at ambient pressure is thought to arise from Mott localization \cite{Gorelov10}, although it has been suggested that disorder-induced localization arising in part from strong spin-orbit coupling may also play a role \cite{Mazin99}.

       \begin{figure} 
            \includegraphics[bb=150 50 600 790,scale=0.76]{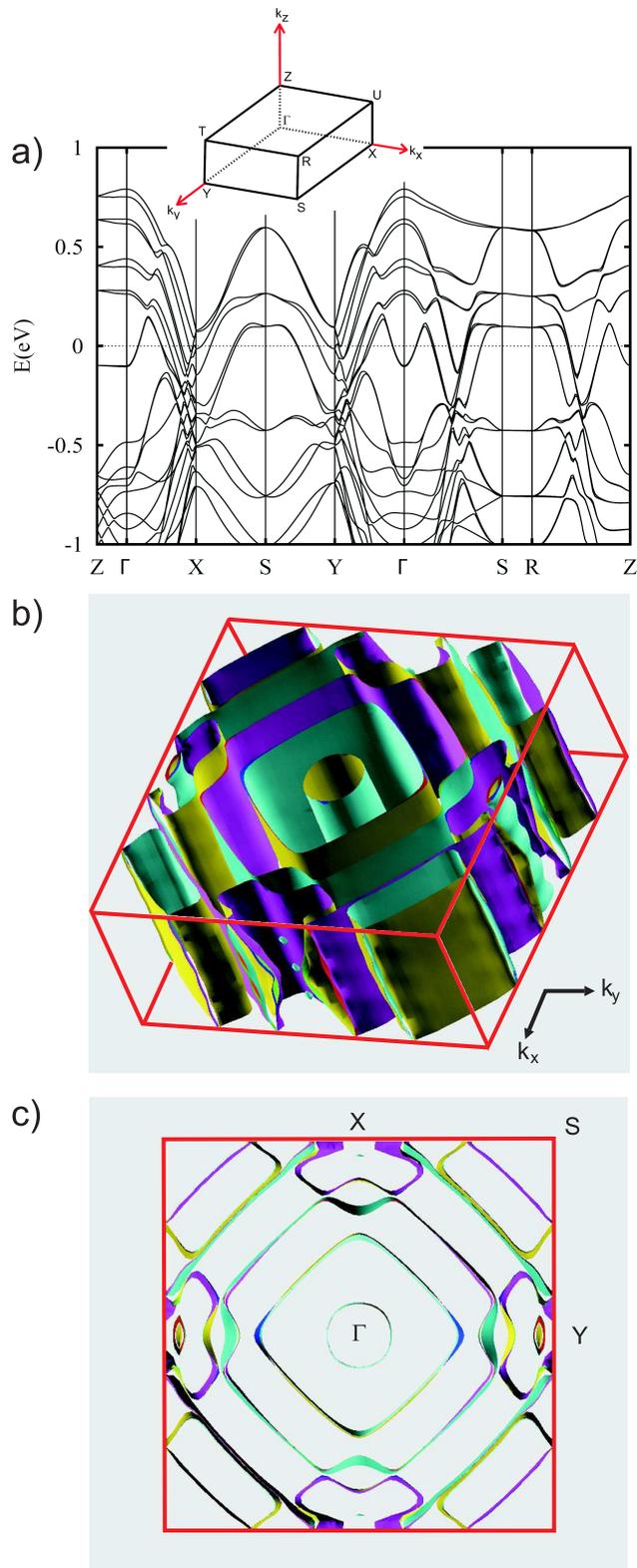}
      \setlength{\abovecaptionskip}{30pt}
              \caption{\label{fig:4} a) Band structure calculations within the LSDA approximation for Ca$_2$RuO$_4$ at a pressure of $p$ = 3.0~GPa. b) the calculated Fermi surface at $p$ = 3.0~GPa, consisting of multiple nested cylinders. c) a slice through the $k_x k_y$ plane.}
              
      \end{figure}
           
In order to interpret our quantum oscillation measurements we calculated the band structure within the local spin density approximation (LSDA) utilizing the general potential linearized augmented plane-wave (LAPW) method. Our calculations were performed using the structural parameters measured by Steffens $et~al.$ \cite{Steffens05} at a pressure of 3~GPa. The effect of spin-orbit coupling has been shown to be very significant in this system \cite{Oguchi09}, and so was included in our calculations. A stable ferromagnetic state with a remnant magnetization of 0.94 $\mu_B$/Ru was obtained, which is higher than the experimental value of 0.4 $\mu_B$/Ru \cite{Nakamura02}. This kind of overestimation of magnetization in standard LSDA calculations is often a sign of renormalisation by quantum spin fluctuations, which is sometimes the case in metals that have competing magnetic ground states and are near critical points. A similar results was seen for instance in the case of iron-pnictide superconductors \cite{Mazin08}.

The predicted band dispersion is shown in Fig. \ref{fig:4}a, along with the full Fermi surface and a cut of the Fermi surface through the basal plane. The two dimensional nature of the material is evident by the lack of dispersion along the S -- R and $\Gamma$ -- Z directions, reflected in the tube-like structure of various sheets of the Fermi surface. The complexity of the predicted Ca$_2$RuO$_4$ Fermi surface compared to the relatively straightforward one observed in isoelectronic Sr$_2$RuO$_4$ \cite{Mackenzie03} is understood to arise from two effects. First, Ca$_2$RuO$_4$ undergoes an orthorhombic reconstruction resulting from rotations of the RuO$_6$ octahedra, which causes band folding with respect to a $\sqrt{2} \times \sqrt{2}$ reconstruction of the Brillouin zone. The ground state is then spin-split due to ferromagnetism, doubling the number of bands that cross the Fermi level, resulting in the multiply nested cylinders seen in Fig. \ref{fig:4}b. At a coarse level, there appears to be a significant number of possible orbits with frequencies in the 3 -- 5~kT range ($\sim$ 20 -- 25 \% of the area of the first Brillouin zone) as well as a number of smaller possible orbits in the $\sim$ 1~kT range.  

Identifying our observed frequencies with orbits around these cylinders however is further complicated by the fact that several of the sheets lie very close together in reciprocal space, such that the conditions for magnetic breakdown are easily achieved in high fields. When the energy gap between two bands $\epsilon_g$ is comparable to the spacing between Landau levels $\hbar\omega_c$, electrons may tunnel from sheet to sheet. This is almost certainly the origin of the 14.6~kT orbit, since an orbit about the entire first Brillouin zone would only yield a frequency of 13.87~kT. However even taking magnetic breakdown effects into account, the very large number of peaks in the oscillation spectrum coupled with the fact that the dHvA frequencies shift between the low field and high field experiments argues against a purely band structure interpretation of our measurements.

       \begin{figure} \centering
             \includegraphics[bb=170 260 400 550,scale=0.63]{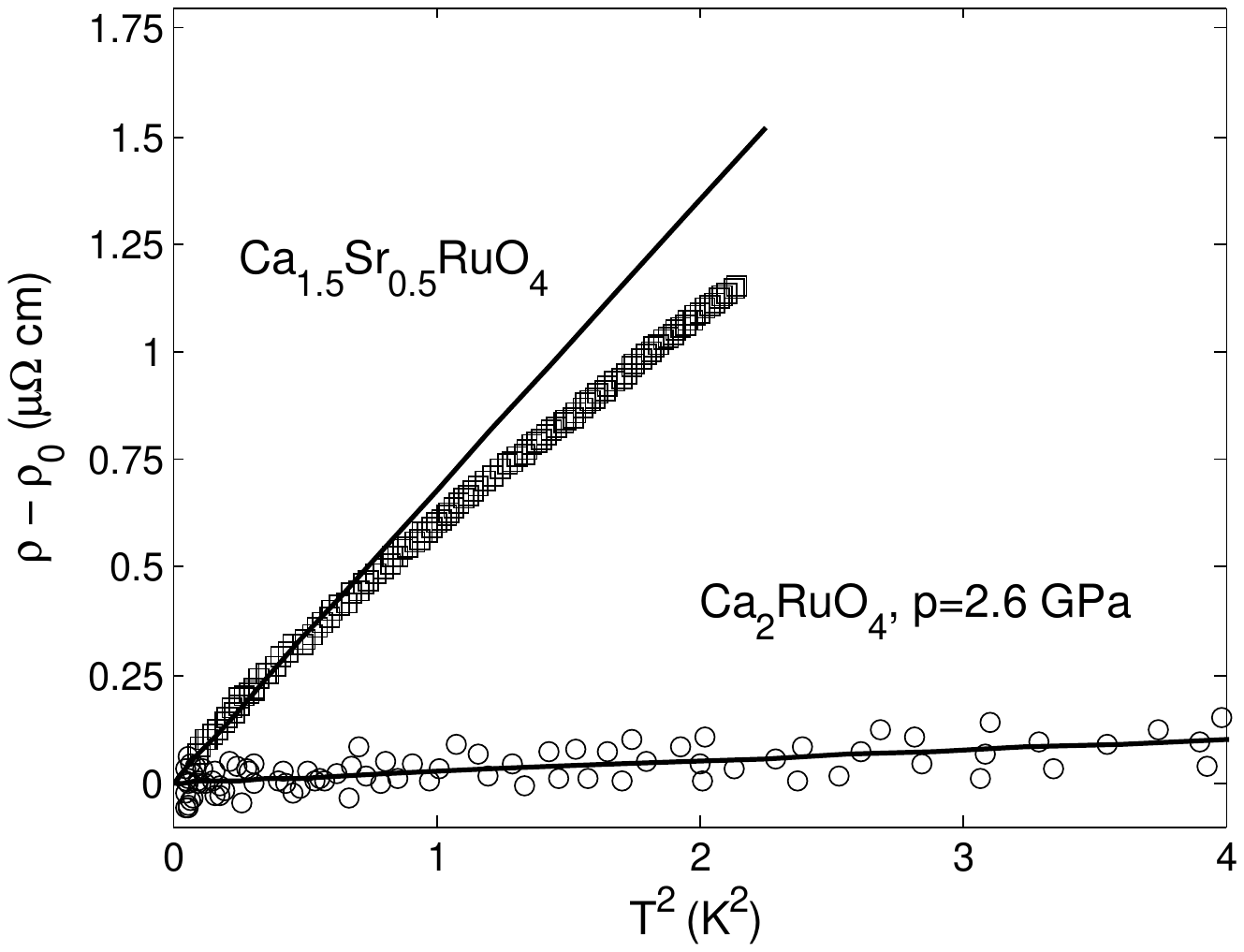}
              \caption{\label{fig:5} In-plane resistivity measurements in Ca$_{2-x}$Sr$_x$RuO$_4$ with $x = 0.5$ at ambient pressure \cite{Nakatsuji03} contrasted with Ca$_2$RuO$_4$ at $p$ = 2.6~GPa \cite{Nakamura02}. The data is taken from the literature, with fits to a $T^2$ temperature dependence shown by solid lines.}
              
      \end{figure}
      
Considering the nature of the band structure, our measurements of light cyclotron masses in Table 1 seem surprising. At $p$ = 2.6~GPa Ca$_2$RuO$_4$ lies between a Mott insulating state at lower pressures, and a critical point separating ferromagnetic order and paramagnetism at higher pressures ($p \gtrsim 7.5$~GPa) \cite{Alireza10}. It is useful to draw an analogy with measurements on the doping system Ca$_{2-x}$Sr$_x$RuO$_4$, where the replacement of Ca with Sr induces chemical pressure. On approach to the Mott transition at $x$ = 0.2, angle resolved photoemission (ARPES) measurements give masses estimated from band dispersions that are very high, $m^*/m_e$ up to 100 in some regions of momentum space  \cite{Shimoyamada09}. At higher dopings, the electronic contribution to the specific heat diverges, reaching a very large value of $\gamma_e$ $\sim$ 255 mJ/mol-RuK$^2$ near the critical point at $x = 0.5$ \cite{Nakatsuji03}. There is strong evidence that this enhancement is driven by the $\gamma$ band, which is flattened by rotations of the RuO$_6$ octahedra \cite{Lee06}. Given that the structure of Ca$_2$RuO$_4$ at $p$ = 2.6~GPa is similar to that of Ca$_{2-x}$Sr$_x$RuO$_4$ near $x = 0.5$, it is reasonable to expect enhanced effective masses in our dHvA experiments as well. 

There are two caveats to keep in mind however. First, the heavy mass Fermi liquid found near $x = 0.5$ in the doping system is realized in the presence of strong magnetic fluctuations, without long range magnetic order. Neutron scattering measurements indicate that overdamped, ferromagnetic-like excitations can quantitatively account for the mass enhancement near the critical region in Ca$_{2-x}$Sr$_x$RuO$_4$ \cite{Friedt04}. However, in the pressure induced metallic state in Ca$_2$RuO$_4$ long range ferromagnetism is present, which would alter the fluctuation spectrum. Measurements of the electrical resistivity confirm that the Fermi liquid states are very different in the two systems, despite the structural similarities. Fig. \ref{fig:5} shows that the $T^2$ coefficient for Ca$_{1.5}$Sr$_{0.5}$RuO$_4$ is some 25 times greater than that of Ca$_2$RuO$_4$ at $p$ = 2.6~GPa, suggesting that the electronic specific coefficient $\gamma_e=C/T$ is at least a factor of five smaller in the pressure induced metal. For a quasi-2D metal, $\gamma_e$ is proportional to the sum of effective masses associated with individual Fermi surface sheets. Given the large number of Fermi surface sheets in our case, it is not unreasonable to measure much smaller effective masses relative to the case of Ca$_{1.5}$Sr$_{0.5}$RuO$_4$.

The second caveat involves the possibility that the masses in Ca$_2$RuO$_4$ are strongly spin dependent, with a heavier majority or minority Fermi surface that remains undetected in our measurements. Masses which are strongly spin dependent are known to arise in scenarios where the Fermi level lies near a peak in the density of states, such as that arising from a van Hove singularity, as is the case here \cite{Fang01}. With these considerations, high pressure inelastic neutron scattering and heat capacity measurements would be useful in developing a complete understanding of the metallic state we observe.

\subsection{Non-linear magnetization effects}

A likely factor contributing to the complexity of the dHvA spectrum in Ca$_2$RuO$_4$ is the itinerant ferromagnetic order present in the metallic state. When a strong magnetic field Zeeman splits the bands crossing the Fermi level, the difference between the up and down-spin Fermi volumes gives the induced magnetization. The size of the Fermi surface thus changes smoothly with field, along with the measured dHvA frequency $f_{obs}$ as \cite{vanRuitenbeck82,Sigfusson84}:

\begin{equation} 
 \label{eq:nlss}
f_{obs} (H) = f(H) - H \frac{df}{dH}+ \ldots
\end{equation}

\noindent where $f$ is the zero field value of the frequency. If the magnetization, and hence $f$ changes linearly with field, the derivative term in Eq. \ref{eq:nlss} can be ignored. In the case of a non-linear field dependent magnetization however, the derivative term causes  $f_{obs}$ to vary with field, with the potential for generating a complex oscillation spectrum. 

       \begin{figure} \centering
            \includegraphics[bb=70 320 550 680,scale=0.52]{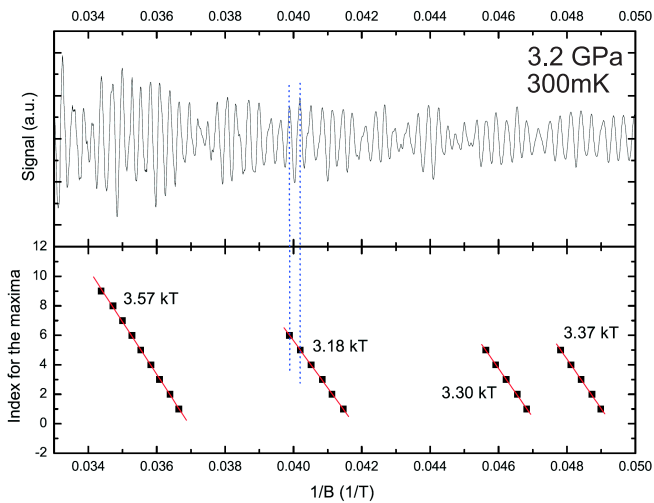}
              \caption{\label{fig:6} Main panel (top): Quantum oscillations in the magnetic susceptibility at $p$ = 3.2~GPa and $T$ = 300~mK, plotted versus $1/B$. Main panel (bottom): Method for determining the oscillation frequency in a narrow field region. The maxima of the oscillations are assigned an index, which is plotted versus $1/B$.}
              
      \end{figure}
      
This phenomenum has been readily observed in band magnets near metamagnetic phase transitions where the magnetization changes abruptly with field, such as in UPt$_3$ \cite{Julian92} and Sr$_3$Ru$_2$O$_7$ \cite{Mercure10}. Although high field magnetization measurements under pressure have not yet been attempted in Ca$_2$RuO$_4$, it is interesting to note that a metamagnetic transition was observed in Ca$_{1.8}$Sr$_{0.2}$RuO$_4$, at B $\sim$ 6~T for B $\parallel$ [001], associated with a destabilization of the antiferromagnetic coupling \cite{Nakatsuji03}. A full magnetic study of the pressure induced ferromagnetic state in Ca$_2$RuO$_4$ would be interesting in this regard.

Further support for this scenario can be found by tracking the measured dHvA frequency $f_{obs}$ as a function of field. The most straightforward way to do this is to employ a simple `counting strategy', where oscillation maxima are used to compute a frequency in a very narrow field range. An example of this analysis for the lowest temperature data at $p$ = 3.2~GPa is shown in Fig. \ref{fig:6}. The raw data is shown in the upper section of the main panel, and the maxima of the oscillations within a narrow region are assigned an index, which is plotted versus $1/B$ in the lower section. The slope of this graph should give a straight line with a slope of $-f_{obs}$, as noted in the figure. What is immediately obvious is that the dominant frequency $-f_{obs}$ is changing markedly with field, from 3.57~kT at the highest fields through a minimum at intermediate fields and to 3.37~kT near $B$ = 20~T. While it is difficult to draw quantitative conclusions from this behavior, the evolution of $-f_{obs}$ with field is at least consistent with the system lying close to a metamagnetic transition at high fields.

Further support for this interpretation would come from magnetization studies of Ca$_2$RuO$_4$ under pressure. It is also possible that extending quantum oscillation measurements to even higher fields would force the material into a fully saturated magnetic state, and a regular dHvA spectrum without complications from non-linear spin splitting effects might be recovered. In this regard, pushing to higher pressures ($p > 7.5$~GPa) would allow access to the paramagnetic metallic state \cite{Alireza10}, which should allow for a more direct test of band structure calculations.

\section{Conclusions}

In summary, our measurements of magnetic susceptibility under pressure in Ca$_2$RuO$_4$ have revealed evidence for quantum oscillations. We interpret these as arising from a homogeneous ferromagnetic metallic state, with a complex oscillatory spectrum which likely arises from strong non-linear magnetization effects. The light cyclotron masses observed in our experiment are considerably less than those seen in similar 2D layered ruthenates, and suggest that heavier Fermi surface sheets may remain to be detected.

\begin{acknowledgments}

We would like to thank S.R. Julian, F.M. Grosche and G.G. Lonzarich for useful discussions. This research was supported by the Royal Society, EPSRC and Trinity College Cambridge. Work at the National High Magnetic Field Laboratory was supported under the auspices of the National Science Foundation and the State of Florida. Work at ORNL was supported by the Department of Energy, Basic Energy Sciences, Materials Sciences and Engineering Division. A part of this work has also been supported by a Grant-in-aid for Scientific Research on Priority Areas from the MEXT of Japan.

\end{acknowledgments}

\bibliography{Ca2RuO4}

\begin{thebibliography}{36}
\bibitem[{end()}]{endnote1}
\bibinfo{note}{In general, it should be noted that the mass estimates for the
  high field work are likely to be slight underestimates of the actual
  cyclotron masses. This is because our fitting method does not allow for the
  fact that the amplitude of the oscillations might be suppressed over a large
  field range due to disorder effects (Eq. 1) as well as thermal smearing.
  Numerical estimates however suggest this is no more than a 25\% effect.}
\end{thebibliography}

\include{Ca2RuO4.bib}

\end{document}